\documentclass[a4paper]{jpconf}
\usepackage{graphicx}
\usepackage{amsmath}
\usepackage{xspace}
\usepackage{upgreek}
\newcommand{\ccbar}{{$\text{c} \overline{\text{c}}$ }\xspace}
\newcommand{\accbar}{{$\langle \text{c} \overline{\text{c}}\rangle$ }\xspace}
\newcommand{\AGeVc}{\ensuremath{A\,\mbox{Ge\kern-0.1em V}\!/\kern-0.08em c}\xspace}
\newcommand{\AGeV}{\ensuremath{A\,\mbox{Ge\kern-0.1em V}}\xspace}

\newcommand{\pip}{\ensuremath{\pi^+}}
\newcommand{\km}{\ensuremath{\textup{K}^-}\xspace}

\newcommand{\Dzero}{\ensuremath{\textup{D}^0}\xspace}
\newcommand{\Dzerobar}{\ensuremath{\overline{\textup{D}^0}}\xspace}
\newcommand{\Dplus}{\ensuremath{\textup{D}^+}\xspace}
\newcommand{\Dminus}{\ensuremath{\textup{D}^-}\xspace}

\begin{document}
\title{Open charm measurements with the NA61/SHINE experiment at the CERN SPS}

\author{Dag Larsen$^1$, Anastasia Merzlaya$^2$ for the NA61/SHINE collaboration}

\address{$^{1,2}$ Jagiellonian University, Krakow, Poland\\$^{2}$ Saint-Petersburg State University, Saint-Petersburg, Russia}

\ead{$^1$ dag.larsen@cern.ch, $^2$ anastasia.merzlaya@cern.ch}

\begin{abstract}
The strong interactions programme of the NA61/SHINE experiment at the CERN SPS has been extended through the use of new silicon Vertex Detector, which allows for reconstructing open charm particles.
The detector was designed to identify those particles by means of a precise reconstruction of their primary and secondary vertices.

An initial version of the Vertex Detector, called SAVD (Small Acceptance Vertex Detector), was installed end of 2016.
It was operated in a Pb+Pb calibration beam time in 2016, and with Xe+La in 2017.
A further Pb+Pb beam time is scheduled for late 2018.
The first indications of a D$^0$ peak were observed in the data obtained from the 2016 Pb+Pb run.

This work introduces the physics motivation behind the open charm measurements, and the status of the analysis of the collected data on open charm production.
Moreover, the plans of future open charm measurements in NA61/SHINE experiment and the related to upgrades of the vertex detector will be discussed.
\end{abstract}

\section{Physics motivation for open charm measurements}
\label{sec1}

\begin{figure}[]
\centering
\includegraphics[width=0.79\linewidth]{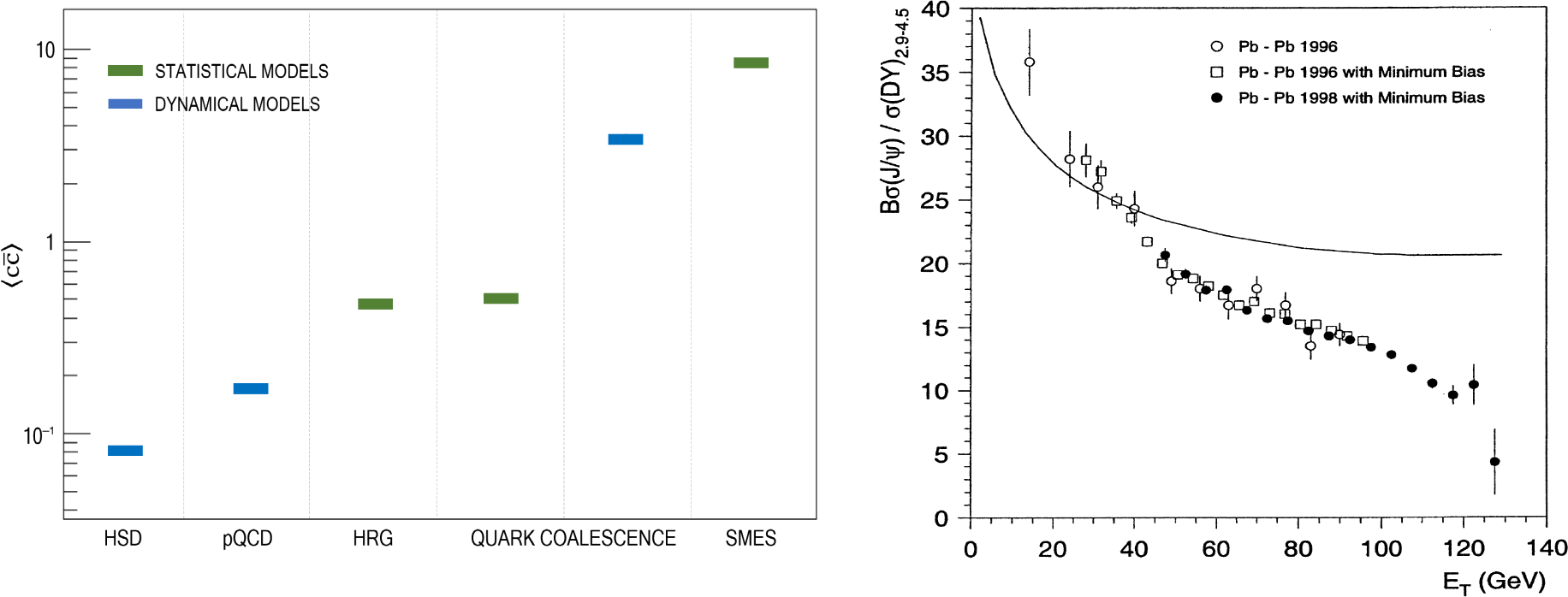}
\caption{
\textit{Left.}
Mean multiplicity of charm quark pairs produced in the full phase space in central Pb+Pb collisions at 158\AGeVc calculated with dynamical models (blue bars): HSD~\cite{Linnyk:2008hp,TSong}, pQCD--inspired~\cite{Gavai:1994gb,BraunMunzinger:2000px}, and Dynamical Quark Coalescence~\cite{Levai:2000ne}, as well as statistical models (green bars): HRG~\cite{Kostyuk:2001zd}, Statistical Quark Coalescence~\cite{Kostyuk:2001zd}, and SMES~\cite{Gazdzicki:1998vd}.
\textit{Right.}
The ratio of $\sigma_{J/\Psi}/\sigma_{DY}$ as a function of transverse energy (a measure of collision violence or centrality) in Pb+Pb collisions at 158 \AGeVc measured by NA50.
The curve represents the J/$\Psi$ suppression due to ordinary nuclear absorption~\cite{Abreu:2000ni}.}
\label{fig:models}
\end{figure}

One of the important issues related to relativistic heavy-ion collisions is the mechanism of charm production.
Several model predictions were introduced to describe charm production.
Some of them are based on the dynamical approach and some on the statistical approach.
The estimates from these approaches for average number of produced $\text{c}$  and $\overline{\text{c}}$ pairs (\accbar) in central Pb+Pb collisions at 158\AGeVc differ by up to a factor of 50 \cite{ana1,Gazdzicki:1998vd} which is illustrated in Figure \ref{fig:models} (\textit{left}).

Charm mesons are of vivid interest in the context of the phase-transition between confined hadronic matter and the quark-gluon plasma (QGP).
The \ccbar pairs produced in the collisions are converted into open charm mesons and charmonia ($\text{J}/\psi$ mesons and its excited states).

The production of charm is expected to be different in confined and deconfined matter.
This is caused by different properties of the charm carriers in these phases.
In QGP, production of charm quark pairs requires 2mC=2.6 GeV.
In hadronic matter, charmed hadrons have to be produced, which requires a higher energy.
A production of a $\rm D\bar D$ pair requires for example 3.7 GeV and thus 1 GeV more than the production of a charm quark pair.
The effective number of degrees of freedom of charm hadrons and charm quarks is similar \cite{Poberezhnyuk:2017ywa}.
Thus, a more abundant charm production is expected in deconfined than in confined matter.
Consequently, in analogy to strangeness \cite{Gazdzicki:1998vd,Rafelski:1982pu}, a change of collision energy dependence of \accbar may be a signal of onset of deconfinement.

Figure~\ref{fig:models} (\textit{right}) shows results on  $\langle \text{J}/\psi \rangle$ production normalised to the mean multiplicity of Drell-Yan pairs in Pb+Pb at the top SPS energy obtained by NA50 collaboration.
The solid line shows model prediction for normal nuclear absorption of $\text{J}/\psi$ in the medium. 
NA50 observed that the $\text{J}/\psi$ production is consistent with normal nuclear matter absorption for peripheral collision and it is suppressed for more central collisions. 
This so called anomalous suppression was attributed to the  $\text{J}/\psi$ dissociation effect in the deconfined medium. 
However, the above result is based on the assumption that \accbar  $\sim \langle DY \rangle $, which may be incorrect due to many effects, such as shadowing or parton energy loss ~\cite{Satz:2014usa}.
Thus, quantitative studies on the \ccbar binding in the medium requires in principle data on the ratio of $\langle \text{J}/\psi \rangle$ to \accbar for A+A and p+p collisions. 
However, the onset of colour screening should already be seen in the centrality dependence of the $\langle \text{J}/\psi \rangle$ to \accbar ratio.
Exploiting this feature allows to avoid separate measurements in p+p collision systems but calls for large statistics data on the \accbar production in A+A collisions.

\section{Performance of SAVD}
\label{sec2}

Estimating \accbar requires to measure the production multiplicities of open charm particles produced in A+A collisions.
To do those measurements, NA61/SHINE was upgraded with a dedicated vertex detector.
The first generation vertex detector, the SAVD is composed from sixteen CMOS MIMOSA-26 sensors~\cite{mimosa26}.
The basic sensor properties are:
1152 columns of 576 pixels with $18.4\times18.4 \rm \upmu m^2$ pitch, $115~\rm \upmu s$ time resolution, $11 \times21~\rm mm^2$ surface, $50~\rm \upmu m$ thickness.	
The estimated material budget per layer, including the mechanical support, is $\approx0.3\%~\rm X_0$.
The sensors were glued to ten ALICE ITS ladders~\cite{Abelev:1625842}, which were mounted on two horizontally movable arms and spaced by 5~cm along the \textit{z} (beam) direction.
The first station is located 5~cm downstream of the target.
The sensors ladders were integrated together with the target into a detector box.
The volume of the box was filled with helium in order to reduce beam-gas interactions.
A first test and calibration run was carried out with the SAVD with a 150 \AGeVc collision system in 2016.
$1.5\times 10^5$ central events were recorded while NA61 was operated in a detector test mode.
The feasibility of tracking in a large track multiplicity environment and an inhomogneous magnetic field was demonstrated~\cite{Merzlaya:2017nvb}.
As track model a parabola in the x-z plane, and a straight line in y-z plane was used~\cite{Merzlaya:2018uta}.

\begin{figure}[]
\centering
\includegraphics[width=0.79\linewidth]{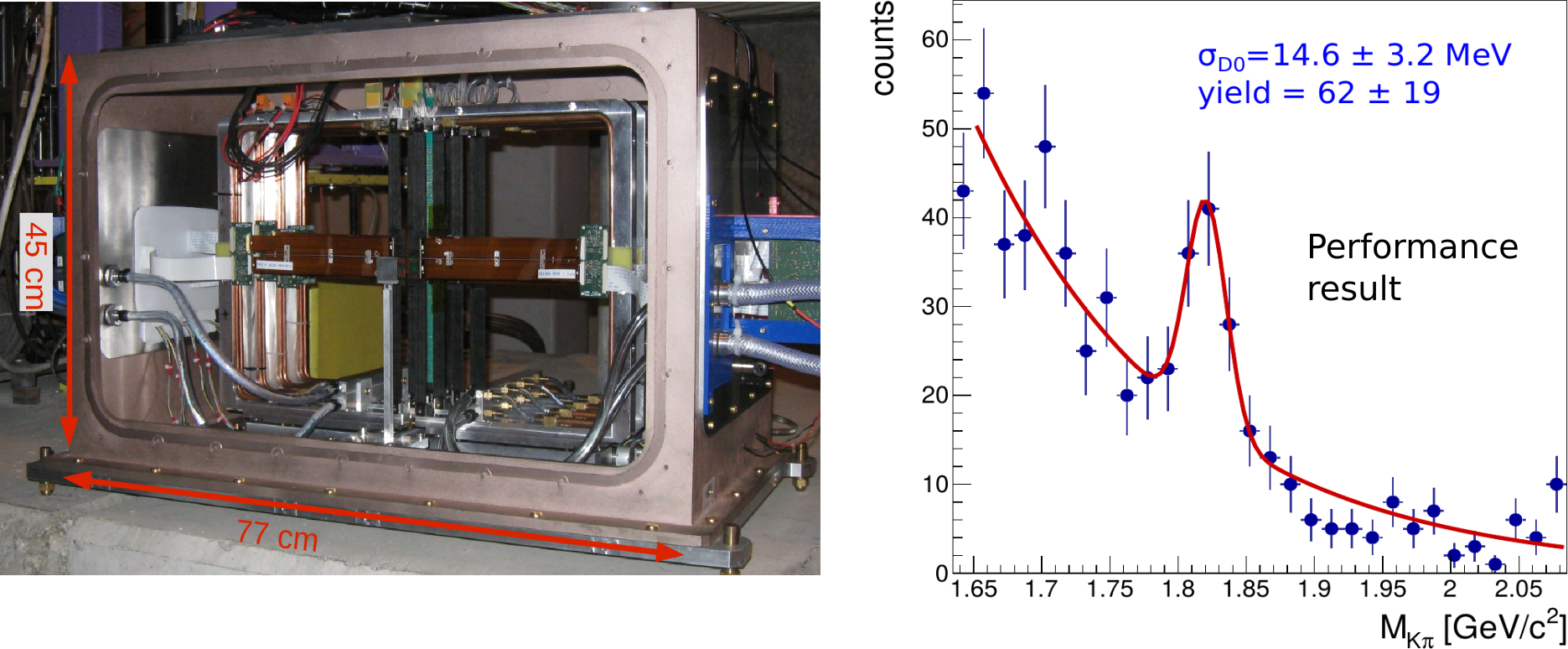}
\caption{
\textit{Left.}
The SAVD used by NA61/SHINE during the data taking in 2016 and 2017. 
\textit{Right.}
The invariant mass distribution of \Dzero and \Dzerobar candidates in central Pb+Pb collisions at 150\AGeVc after the background suppression cuts.
The particle identification capability of NA61/SHINE was not used at this stage of the analysis~\cite{ana1}.
}
\label{fig:firstD0}
\end{figure}

Based on these data, the spatial resolution of the SAVD was determined. 
The cluster position resolution is $\sigma_{\text{x,y}}(\text{Cl}) \approx 5~\rm \upmu m$.
The primary  vertex resolution in the transverse plane is $\sigma_{\text{x}}(\mathrm{PV}) \approx  5~\rm \upmu m $, $\sigma_{\text{y}}(\text{PV}) \approx 1.8~\rm \upmu m $\footnote{$\sigma_{\text{x}}(\text{PV}) > \sigma_{\text{y}}(\text{PV})$; because $B_\text{y} \gg B_\text{x}$ a more precise straight-line fit is used for y-z, while a parabolic fit has to be used in x-z.}in the transverse plane and $\sigma_{\text{z}}(\text{PV}) \approx 30~\rm \upmu m$ along the beam direction for a typical multiplicity of events recorded in 2016. 
This resolution was sufficient to perform the search for the \Dzero and \Dzerobar signals.
Figure~\ref{fig:firstD0} (\textit{right}) shows the first indication of a \Dzero and \Dzerobar peak obtained using the data collected during the Pb+Pb run in 2016.

\begin{figure}[]
\centering 
\begin{minipage}{0.60\linewidth}
\includegraphics[width=\linewidth]{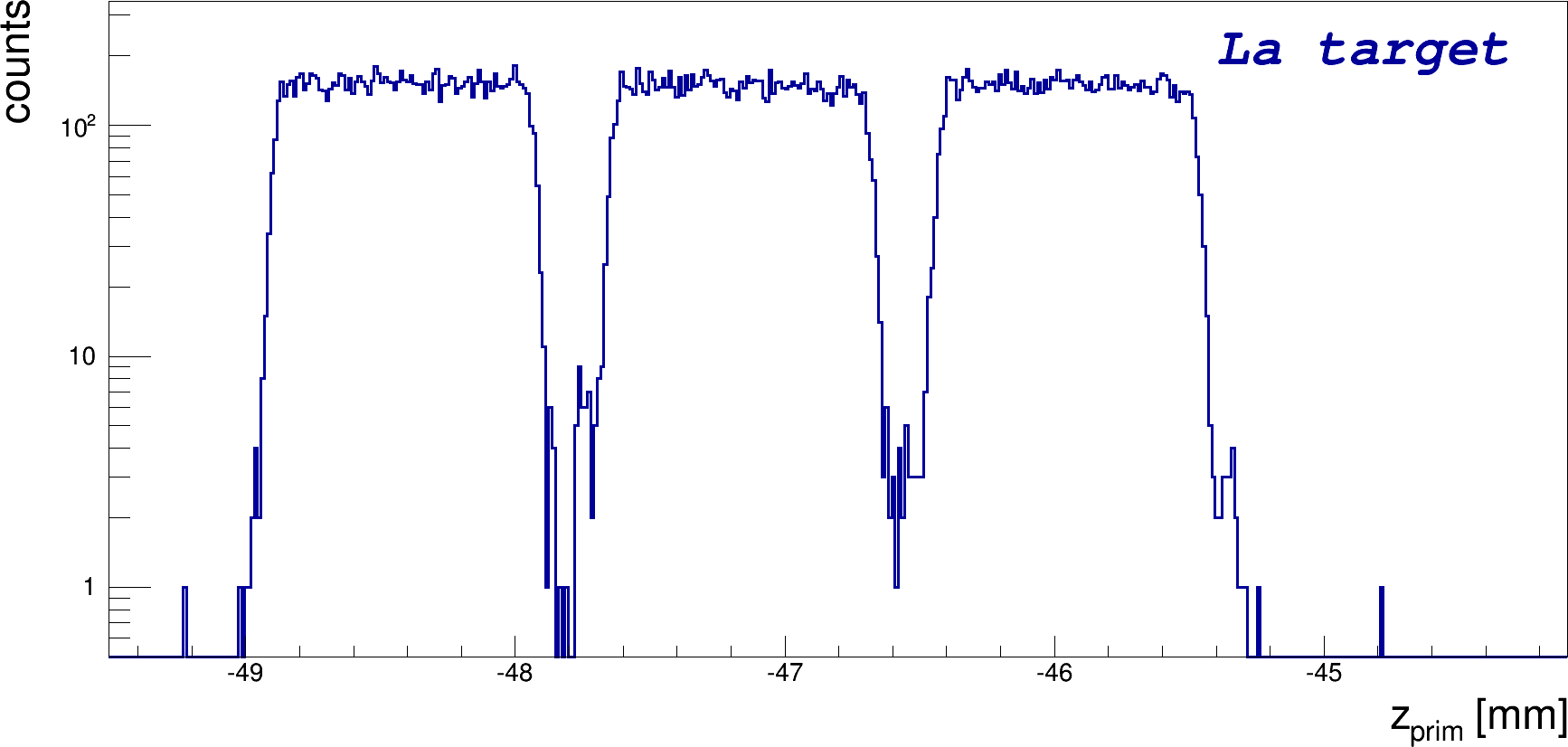} 
\end{minipage}
\hspace{0.04\linewidth}
\begin{minipage}{0.34\linewidth}
\caption{Distribution of longitudinal coordinate of the primary vertex $z_{prim}$ for interactions in the La target, which was composed of three 1~mm plates.} 
\label{fig:vd-prim-longitu} 
\end{minipage}
\end{figure}

The good performance of the SAVD in 2016 led to the decision to also use it during the Xe+La data taking in 2017.
About $5\cdot10^6$ events of central Xe+La collisions at 150\AGeVc were collected in October and November 2017.
In the preparation of the beam time, the threshold tuning procedure was improved, which led to a substantial improvement of the detection efficiency of the LAVD.
This and a reworked alignment allowed for improving the vertex resolution for Xe+La data in the transverse and longitudinal coordinates to $1~\rm \upmu m$ and $15~\rm \upmu m$ respectively.
The distribution of the longitudinal coordinate ($z_{prim}$) of the primary vertex is shown in Figure \ref{fig:vd-prim-longitu} (\textit{left}) (see ~\cite{ana1} for details).
The Xe+La data are currently under analysis and are expected to lead to physics results in the coming months.

The SAVD will also be used during a scheduled, three weeks long Pb+Pb data taking at 150\AGeVc in 2018.
It is foreseen to record $\approx 10^7$ central collisions, which would presumably allow to reconstruct 2500 \Dzero + \Dzerobar.

\section{Proposed measurements after Long Shutdown 2}
\label{sec3}


During the Long Shutdown 2 (LS2) at CERN (2019-2020), a significant modification of the NA61/SHINE spectrometer is planned.
The upgrade is primarily motivated by the charm program which requires a ten fold increase of the data taking rate to about 1~kHz.
Also, an increase of the phase-space coverage of the VD by a factor of about 2 is planned.
This requires an upgrade of the VD, replacement of the TPC read-out electronics, implementation of new trigger and data acquisition systems and upgrade of the Projectile Spectator Detector.
Finally, new ToF detectors are planned to be constructed for particle identification at mid-rapidity.

The detector upgrades are discussed in detail in \cite{ana1}.
The data taking plan related to the open charm measurements covers measurements of 500M inelastic Pb+Pb collisions at 150\AGeVc in 2021--2024.
This data will provide the mean number of \ccbar pairs in central Pb+Pb collisions needed to investigate the mechanism of charm production in this reaction.
Moreover, the data will allow to establish the centrality dependence of \accbar in Pb+Pb collisions at 150\AGeVc and thus address the question of how the formation of QGP impacts $ \text{J}/\psi $ production.  
Table~\ref{tab:centrality} lists the expected number of reconstructed charm mesons in per centrality bin for Pb+Pb collisions at 150\AGeVc assuming the mentioned above statistics of minimum bias collisions.
The estimate is performed assuming that mean multiplicity of charm hadrons is proportional to the number of binary collisions and using yields calculated for central Pb+Pb collisions within the HSD model~\cite{Linnyk:2008hp,TSong}.
Central (0-30\%) Pb+Pb collisions at 40\AGeVc are planned to be recorded in 2024.
This data together with the result for central Pb+Pb collisions at 150\AGeVc will start a long-term effort to establish the collision energy dependence of \accbar and address the question of how the onset of deconfinement impacts charm production.

\begin{table}[]
\centering
\begin{tabular}{c c c c c c c }
&   0--10\%  & 10--20\% & 20--30\%  &  30--60\%   &  60--90\%  &  0--90\%   \\ \hline 
\#(\Dzero+ \Dzerobar)      &    31k    &   20k     &    11k    &  13k        &   1.3k       &  76k      \\
\#(\Dplus+ \Dminus)        &    19k    &   12k     &    7k    &    8k        &   0.8k     &    46k      \\
$\langle W \rangle$        &    327    &   226     &   156     &  70         &   11       &  105       \\
$\langle N_{COLL} \rangle$   &    749    &   499     &           & 102         &   11       &  202       \\
\end{tabular}
\caption{
Number of reconstructed D-mesons as function of centrality for 500 M minimum bias events for a Pb+Pb collision system at 150 \AGeVc.
$\langle W \rangle$ and $\langle N_{COLL} \rangle$ denote the mean number of wounded nucleons and the mean number of binary collisions respectively.
}
\label{tab:centrality}
\end{table}
The expected high statistics of reconstructed \Dzero and \Dzerobar decays is due to the high event rate and the relatively large efficiencies of open charm detection in the VD. 
The efficiency will be about 13\% (3 times better than for the SAVD) for the \Dzero  $\rightarrow \pip + \km$ decay channel and about 9\% \footnote{The quoted efficiencies include the geometrical acceptance for \Dzero $\rightarrow \pip + \km$ (\Dplus $\rightarrow \pip + \pip + \km$) decays and the efficiency of the analysis quality cuts used to reduced the combinatorial background.} for \Dplus decaying into $\pip + \pip + \km$. 

\begin{figure}[]
\centering
\begin{minipage}{0.34\linewidth}
\caption{
Transverse momentum and rapidity distribution of \Dzero+ \Dzerobar mesons produced in about 500M inelastic Pb+Pb collisions at 150\AGeVc for all produced \Dzero+ \Dzerobar mesons (\textit{left}) and  \Dzero+ \Dzerobar mesons reconstructed in VD and  passing background suppression cuts \textit{(right)}.}
\label{fig:acceptSimLAVD}
\end{minipage}
\hspace{0.04\linewidth}
\begin{minipage}{0.60\linewidth}
\includegraphics[width=\linewidth]{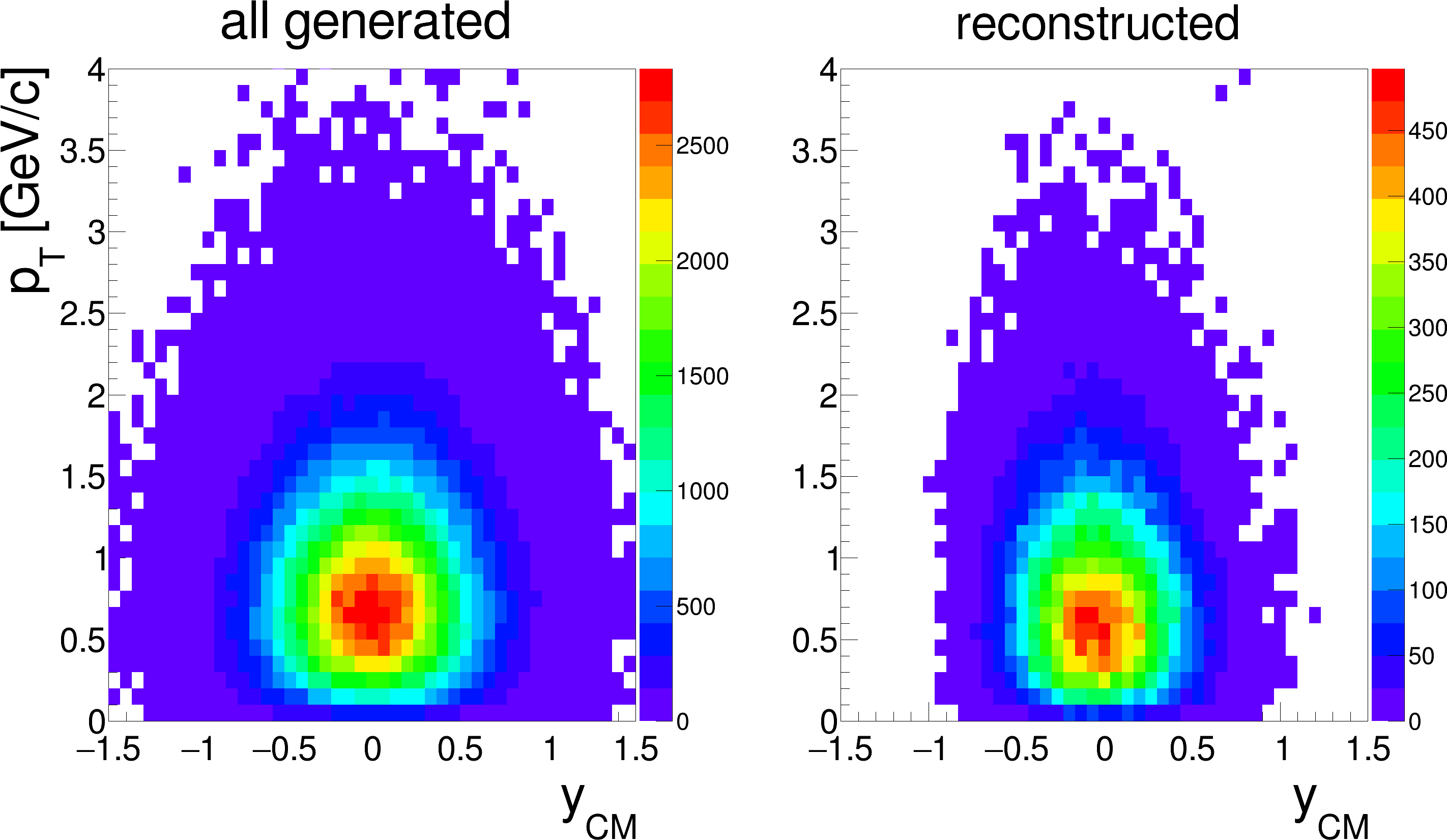}
\end{minipage}
\end{figure}

\begin{figure}[]
\centering
\begin{minipage}{0.60\linewidth}
\includegraphics[width=\linewidth]{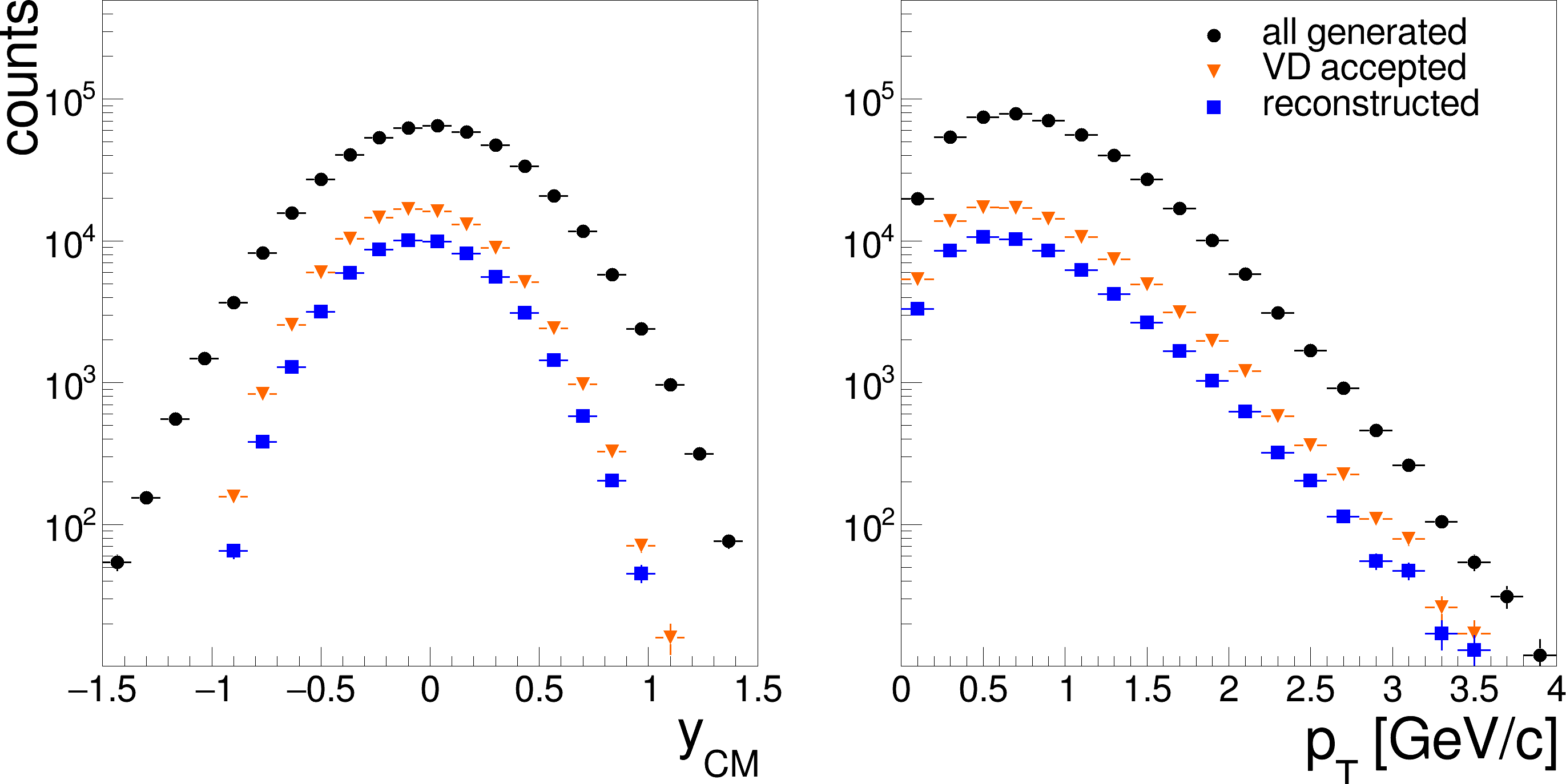}
\end{minipage}
\hspace{0.04\linewidth}
\begin{minipage}{0.34\linewidth}
\caption{
Rapidity (\textit{left}) and  transverse momentum (\textit{right}) distributions of \Dzero+ \Dzerobar mesons produced in about 500M inelastic Pb+Pb collisions at 150\AGeVc.
}
\label{fig:y:full}
\end{minipage}
\end{figure}

Figure~\ref{fig:y:full} shows distributions of \Dzero + \Dzerobar mesons in rapidity and transverse momentum for all generated particles (black symbols)  and for particles that passed the acceptance and background reduction cuts (blue symbols).
The presented plots refer to 500M inelastic Pb+Pb collisions at 150\AGeVc. 
Based on the presented simulations one estimates, that fully corrected results will correspond to more than 90\% of the \Dzero and \Dzerobar yield (see Figures~\ref{fig:acceptSimLAVD} and~\ref{fig:y:full}).
The total uncertainty of $ \langle \Dzero \rangle $ and $ \langle \Dzerobar \rangle $ is expected to be about 10\% and dominated by systematic uncertainties.

\section{Summary}
\label{sec4}
Only NA61/SHINE is able to measure open charm production in heavy ion collisions in full phase space at SPS energies. 
The corresponding potential measurements at higher (LHC, RHIC) and lower (FAIR, J-PARC) energies are necessary to complement the NA61/SHINE results and establish collision energy dependence of charm production.
The open charm programme of NA61/SHINE is on a good track and unique results are expected.

\section*{Acknowledgements}
This work was supported by the Polish National Centre for Science grants 2014/15/B/ST2/02537 and 2015/18/M/ST2/00125.

\end{document}